\documentclass[journal]{IEEEtran}
\usepackage[english]{babel}
\usepackage{amsmath}
\usepackage{graphicx}
\usepackage[colorlinks=true, allcolors=blue]{hyperref}
\usepackage{xcolor}
\usepackage{comment}
\usepackage{lipsum}
\usepackage{lettrine}
\usepackage{cite}  
\usepackage{multicol}
\usepackage{makecell}
\usepackage{colortbl}
\usepackage{xcolor}
\usepackage{booktabs}
\usepackage{balance}
\usepackage{csquotes}

\newcommand{\orcidiconBeo}{\href{https://orcid.org/0000-0001-5929-1672}{\includegraphics[scale=0.1]{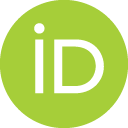}}}

\newcommand{\orcidiconOba}{\href{https://orcid.org/0000-0003-2523-3858}{\includegraphics[scale=0.1]{figures/orcidID128.png}}}

\begin{document}
\bstctlcite{IEEEexample:BSTcontrol}
\setlength{\parskip}{0pt}

\title{Modeling and Analysis of SCFA-Driven Vagus Nerve Signaling in the Gut-Brain Axis via Molecular Communication}

\author{Beyza E. Ortlek\orcidiconBeo,~\IEEEmembership{Student Member,~IEEE}, and~Ozgur~B.~Akan\orcidiconOba,~\IEEEmembership{Fellow,~IEEE}
\thanks{Beyza E. Ortlek is with the Center for neXt-generation Communications (CXC), Department of Electrical and Electronics Engineering, Ko\c{c} University, Istanbul 34450, Turkey (e-mail: bortlek14@ku.edu.tr).}
\thanks{Ozgur B. Akan is with the Center for neXt-generation Communications (CXC), Department of Electrical and Electronics Engineering, Ko\c{c} University, Istanbul 34450, Turkey and also with the Internet of Everything (IoE) Group, Electrical Engineering Division, Department of Engineering, University of Cambridge, Cambridge CB3 0FA, UK (email: oba21@cam.ac.uk).}
\thanks{This work was supported in part by the AXA Research Fund (AXA Chair for Internet of Everything at Ko\c{c} University).}
}

\maketitle

\begin{abstract}
Molecular communication (MC) is a bio-inspired communication paradigm that utilizes molecules to transfer information and offers a robust framework for understanding biological signaling systems. This paper introduces a novel end-to-end MC framework for short-chain fatty acid (SCFA)-driven vagus nerve signaling within the gut-brain axis (GBA) to enhance our understanding of gut-brain communication mechanisms. SCFA molecules, produced by gut microbiota, serve as important biomarkers in physiological and psychological processes, including neurodegenerative and mental health disorders. The developed end-to-end model integrates SCFA binding to vagal afferent fibers, G protein-coupled receptor (GPCR)-mediated calcium signaling, and Hodgkin-Huxley-based action potential generation into a comprehensive vagus nerve signaling mechanism through GBA. Information-theoretic metrics such as mutual information and delay are used to evaluate the efficiency of this SCFA-driven signaling pathway model. Simulations demonstrate how molecular inputs translate into neural outputs, highlighting critical aspects that govern gut-brain communication. In this work, the integration of SCFA-driven signaling into the MC framework provides a novel perspective on gut-brain communication and paves the way for the development of innovative therapeutic advancements targeting neurological and psychiatric disorders.
\end{abstract}


\section{Introduction}

\IEEEPARstart{M}{olecular} communication (MC) is a bio-inspired communication technique that aims to mimic biological and chemical communication mechanisms and provide a theoretical framework for molecular-level processes from an information and communication technology (ICT) perspective \cite{akan2016fundamentals,farsad2016comprehensive}. Understanding molecular signaling processes existing in nature via MC enables envisioning the Internet of Bio-Nano Things (IoBNT) technology \cite{kuscu2021internet}. This IoBNT technology employs seamless communication between the natural and artificial nano-biological functional devices integrated into the Internet infrastructure  \cite{akyildiz2010internet,akyildiz2015internet}. Revealing the fundamentals of molecular nanonetworks from an ICT perspective also enables many of the foreseen IoBNT applications that further provide novel and effective ICT-based diagnosis and treatment techniques for human diseases.

\begin{figure}[h]
	\centering
	\includegraphics[width=1\columnwidth]{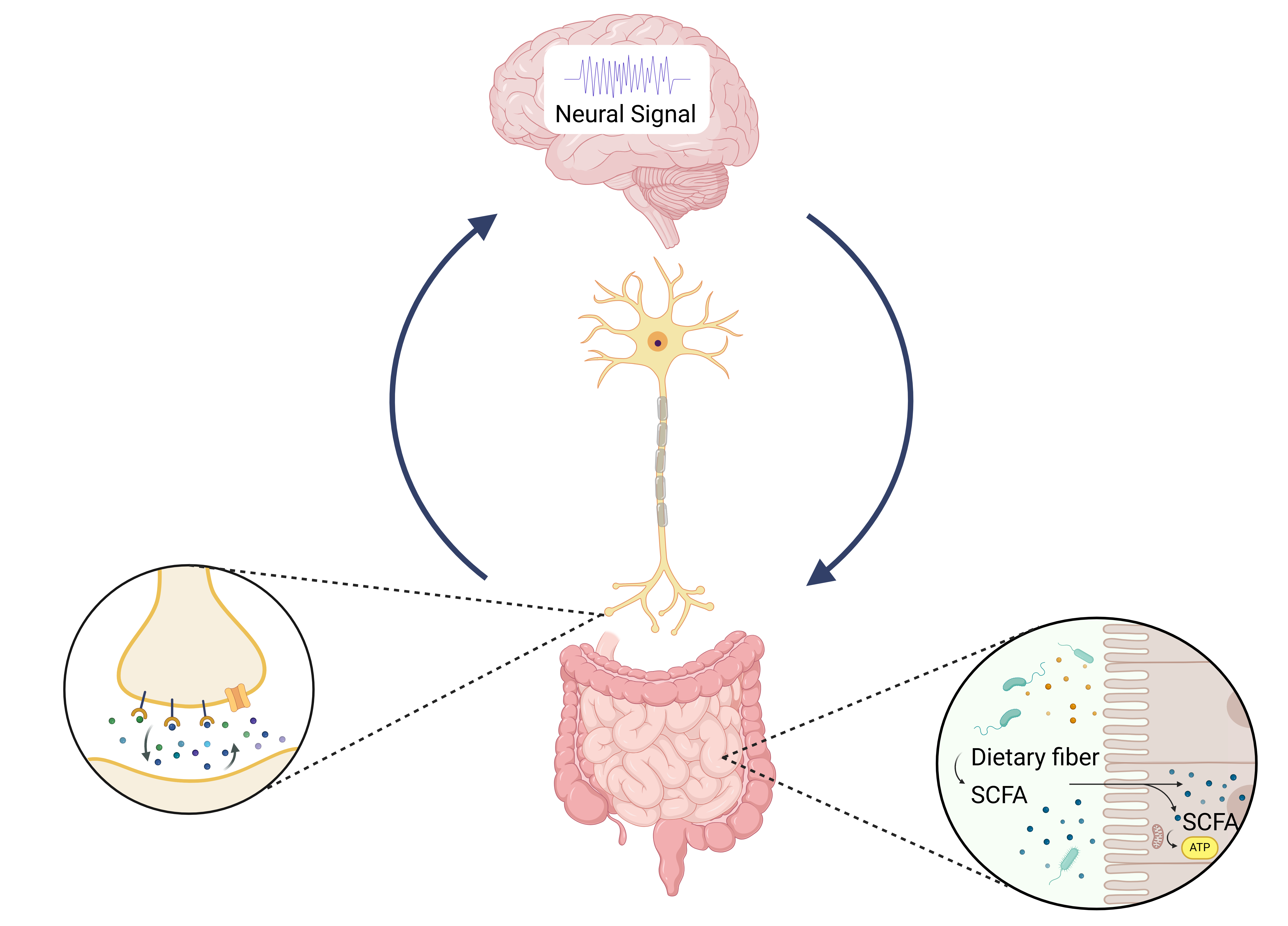}
	\caption{Biological Representation of MC System (Created in  https://BioRender.com).}
	\label{fig:vagal_illustration}
\end{figure}

The gut-brain axis (GBA) is a bidirectional communication network that facilitates interactions between the gastrointestinal (GI) tract and the central nervous system (CNS), as depicted in Fig. \ref{fig:vagal_illustration}. GBA is a complex communication network that includes neural, endocrine, immune, and humoral pathways and plays a pivotal role in regulating numerous metabolic and neuropsychological processes \cite{mayer2011gut}. Recent advancements have underscored the significant role of gut microbiota and their metabolic by-products, particularly short-chain fatty acids (SCFAs) such as butyrate, acetate, and propionate, in modulating the GBA. SCFAs, produced by gut bacteria through the fermentation of dietary fibers, are integral to various signaling pathways within the GBA, encompassing neural, immune, and endocrine routes \cite{mayer2022gut,silva2020role,dalile2019role,ortlek2023communication}.

The vagus nerve (\textit{cranial nerve X}) plays a central role in the GBA by transmitting a wide array of signals between the digestive system and the brain. Originating in the medulla oblongata, the vagus nerve extends through the neck, thorax, and abdomen, earning the nickname \enquote{wanderer nerve} due to its extensive path through the body \cite{breit2018vagus,rosas2011acetylcholine}. It innervates most of the pharynx and larynx muscles, which are essential for swallowing and vocalization, provides the primary parasympathetic innervation to the heart, and regulates smooth muscle contractions and glandular secretions in the intestines. Abdominal vagal afferents, including mechanoreceptors and chemoreceptors, detect SCFAs in the gut lumen and transmit these signals to the nucleus tractus solitarius (NTS) in the brainstem, integrating gut-derived information with higher-order brain functions \cite{berthoud2000functional,carabotti2015gut}.

In the neural pathway, SCFAs bind to G-protein-coupled receptors (GPR41/FFAR3) on vagal afferent neurons, initiating neuronal signals that propagate along the vagus nerve to the brain. This binding activates intracellular signaling cascades, leading to membrane depolarization, action potential generation, and the transmission of neural signals from the gut to the NTS in the brainstem\cite{bonaz2018vagus,dalile2019role,longo2023microbiota}. Concurrently, SCFAs modulate cytokine production and immune cell activity in the immune pathway, thereby influencing inflammatory responses and immune homeostasis. In the endocrine pathway, SCFAs regulate hormone secretion and metabolic processes, impacting stress responses, mood, and appetite regulation \cite{margolis2021microbiota, cryan2019microbiota}.

The production of SCFAs is a critical biochemical process occurring primarily in the colon, where specific gut bacteria ferment non-digestible carbohydrates such as dietary fibers and resistant starches. The predominant SCFAs produced are acetate (60-70\%), propionate (20-25\%), and butyrate (10-15\%), each playing distinct roles in maintaining gut health and systemic physiological balance. Beyond serving as energy substrates for colonocytes, SCFAs act as signaling molecules that travel the GBA, mediating communication between the gut microbiota and the CNS. This intricate interplay underscores the significance of SCFAs in local gut homeostasis and broader neurological functions \cite{o2022short,guo2022gut}.

Dysregulation of SCFA production and signaling has been implicated in a spectrum of diseases and health conditions. Inflammatory bowel diseases (IBD), including Crohn's disease and ulcerative colitis, are characterized by altered SCFA levels, which contribute to impaired gut barrier integrity and chronic inflammation. Moreover, diminished butyrate production has been associated with colorectal cancer, given its anti-proliferative and pro-apoptotic properties in colonocytes. Beyond gastrointestinal disorders, SCFAs influence metabolic diseases such as obesity and type 2 diabetes by modulating insulin sensitivity and lipid metabolism. Imbalances in SCFA-mediated signaling within the GBA are linked to neurological and psychiatric disorders, including depression, anxiety, and autism spectrum disorders (ASD). SCFAs impact brain function by modulating neurotransmitter synthesis, neuroinflammation, and neuroplasticity, highlighting their critical role in mental health \cite{majumdar2023short, xiao2022microbiota, o2022short, silva2020role}.

Integrating MC principles into studying SCFA-mediated signaling within the GBA offers a unique opportunity to quantify and optimize information transfer and signaling capacities inherent in these biological communication networks. This approach enhances our understanding of physiological communication mechanisms and informs the design of IoBNT-enabled diagnostic and therapeutic devices. Such devices can be engineered to modulate SCFA levels and signaling pathways, providing targeted interventions for related diseases.

Exploring the GBA through the perspective of MC is a relatively new and developing field, with only a few pioneering studies addressing the complex communication mechanisms involved. Building a robust research framework for the IoBNT within this context, \cite{akyildiz2019microbiome} presents the Microbiome-Gut-Brain Axis (MGBA) as a biomolecular communication network. This study addresses fundamental challenges in developing a self-sustainable and biocompatible network infrastructure to interconnect next-generation wearable and implantable devices. By investigating minimally invasive and externally accessible electrical and molecular communication channels, the research explores how information can be transmitted through the MBGA. The proposed comprehensive framework integrates with the biological processes of the MGBA and intercomponent communications, aiming to overlay a network infrastructure onto these processes. These initial investigations have begun to elucidate how MC can be employed to model and understand the intricate signaling pathways between the gut and the brain.

In our previous study \cite{ortlek2023communication}, we delved into the foundational aspects of MC within the GBA, establishing a basis for understanding how molecular signals regulate complex physiological interactions. We developed an MC-based model specifically for p-cresol, providing a comprehensive analysis of its propagation mechanisms within the GBA. By characterizing the impulse response of the MC channel and performing extensive numerical simulations, valuable insights are obtained into how variations in system parameters, such as transmitter-receiver distances and receiver dimensions, influence p-cresol signaling in ASD. These findings enhance understanding of the molecular mechanisms underlying ASD and demonstrate the potential of applying communication theoretical frameworks to unravel biological processes through the GBA.

Expanding the theoretical foundations of MC in the GBA, \cite{maitra2024molecular} introduces the concept of a Molecular Quantum (MolQ) communication channel at the synapse of the GBAx. This approach focuses on the diffusion of neurotransmitters, such as acetylcholine, through the synaptic cleft to the vagus nerve membrane (VNM), forming ligand-receptor complexes that trigger ion channel openings. By integrating Quantum Communication (QC) with MC, the MolQ communication model provides a comprehensive theoretical framework that elucidates the entire signaling pathway from neurotransmitter release to ion channel activation, offering novel insights into the molecular mechanisms governing the GBA and their implications for neurological health.

These pioneering studies collectively establish the foundation for understanding the GBA through MC frameworks, demonstrating the feasibility and potential benefits of applying advanced communication theories to biological systems. Despite these advancements, the field remains in its infancy, with many biological signaling mechanisms within the GBA yet to be fully elucidated.

To address this, this paper introduces a comprehensive model for SCFA-mediated vagus nerve signaling within the GBA and, to the best of our knowledge, represents the first study to examine this pathway from an MC perspective. By delving deeper into the neural pathways and employing MC principles, the study models and analyzes the efficiency and reliability of SCFA-induced neuronal signaling, providing insights into the underlying communication mechanisms that influence neurological and psychiatric health. By conceptualizing this pathway as a communication channel, the study aims to:

\begin{itemize}
    \item Model SCFA-dependent receptor binding kinetics, capturing how ligand concentration influences receptor activation and neuronal response.
    \item Describe the action potential generation and propagation process along the vagus nerve, including noise factors that may influence signal fidelity.
    \item Analyze neurotransmitter release dynamics in the nucleus tractus solitarius (NTS), linking the firing rate from vagal afferents to downstream signaling in the brain.
    \item Conduct an information-theoretic analysis of the pathway, evaluating metrics such as mutual information and delay resilience.
\end{itemize}

This paper contributes to the ongoing efforts to understand and optimize MC pathways within the GBA, offering new perspectives on integrating communication theory with biological processes. The rest of the paper is organized as follows. Section II introduces the system model, detailing the SCFA-driven vagus nerve signaling mechanisms within the GBA. Section III presents the theoretical analysis and simulations, where the proposed MC framework is evaluated using information-theoretic metrics such as mutual information and delay. Section IV discusses the simulation results, highlighting the impact of SCFA-induced G-protein activation rates on communication efficiency and signaling delays. Finally, Section V concludes the paper and suggests directions for future research.

\section{System Model}

The GBA is a complex, bidirectional communication network linking the GI tract to the CNS through neural, endocrine, and immune pathways. One of the central components of this network is the \textit{vagus nerve}, i.e., the tenth cranial nerve (CN X), which provides a direct neural connection between the brainstem and various peripheral organs, including the heart, lungs, and GI tract \cite{berthoud2000functional}. Originating in the brainstem, the vagus nerve extends through the neck and thorax to reach the abdominal organs. Approximately 80\% of the vagal fibers are afferent, transmitting sensory information from the peripheral organs to the CNS, while the remaining 20\% are efferent, conveying motor signals that modulate physiological functions \cite{han2022vagus}. Within the GBA, these afferent fibers play a crucial role in monitoring the internal environment of the gut and conveying changes to the CNS. This sensory feedback helps maintain homeostasis and influences mood, behavior, immune responses, and autonomic regulation \cite{bonaz2018vagus}.

SCFAs, including acetate, propionate, and butyrate, are important signaling molecules involved in gut-to-brain communication. These molecules are produced by the gut microbiota through the fermentation of dietary fibers and serve as key metabolic outputs that reflect the microbial composition and activity within the GI tract. SCFA molecules interact with G protein-coupled receptors (GPCRs), particularly FFAR3 (formerly GPR41), located on the terminals of vagal afferent neurons in the nodose ganglion\cite{husted2017gpcr}. By binding to these receptors, SCFAs initiate complex intracellular signaling cascades that ultimately modulate neuronal excitability and action potential generation. SCFAs act as molecular messengers that encode information about the microbial environment of the gut and transmit it to the CNS via the vagus nerve \cite{ikeda2022short,dalile2019role,silva2020role}.

The SCFA-driven calcium signaling cascade refines both the timing and accuracy of neuronal responses, ensuring that vagal afferent fibers generate action potentials that reliably represent gut microbial activity and the gastrointestinal environment. As these action potentials propagate along the vagus nerve, they provide the CNS with timely, precise updates, enabling adjustments in mood regulation, stress responses, and autonomic functions in response to changing gut conditions. Moreover, the intricate interplay of SCFA binding, G protein activation, and calcium signaling preserves the temporal responsiveness of neuronal communication within the GBA. This efficient transfer of information ensures that the CNS can promptly and accurately interpret gut-derived signals, thereby supporting coordinated physiological and behavioral adaptations.

This advanced MC mechanism highlights the importance of SCFAs in maintaining homeostasis and influencing neurophysiological functions through the GBA. Understanding these molecular mechanisms provides valuable insights into how gut-derived signals can impact brain function and behavior, paving the way for potential therapeutic interventions targeting neurological and psychiatric disorders.

\subsection{SCFA-GPCR Binding and Intracellular Signaling}

Within the MC framework, the vagal pathway is effectively modeled by considering the concentration of SCFA molecules as the primary input signal. This signal is transmitted through receptor-ligand binding and ultimately converted into electrical signals. The vagal pathway can be conceptualized as a sequence of processes within the MC framework, encompassing the following components:

\begin{itemize}
	\item \textbf{Input Signal}: Gut microbiota-derived SCFA molecules serve as the initial input signal, representing the molecular information transmitted within the GBA channel.
	\item \textbf{Channel and Transmission Mechanism}: The SCFAs diffuse across the gut environment and bind to G protein-coupled receptors (FFAR3/GPR41) on vagal afferent neurons. This binding activates the GPCR complex and initiates intracellular signaling cascades that lead to the generation of action potentials.
	\item \textbf{Output Signal}: The action potentials propagate along the vagus nerve to the NTS in the brainstem, resulting in the release of neurotransmitters. These neurotransmitters modulate higher brain functions and complete the communication pathway.
\end{itemize}

The signal processing within the vagal pathway can be divided into three major stages:
\begin{enumerate}
	\item  \textbf{Diffusion and Binding}: SCFAs produced by gut microbiota diffuse across the GI lining to interact with receptors on vagal afferent neurons located in the nodose ganglion.
	\item \textbf{Intercellular Signaling, Action Potential Generation and Propagation}: The binding of SCFAs to FFAR3 receptors triggers a series of intracellular events that result in membrane depolarization. This depolarization initiates action potentials, which encode the molecular signal and propagate along the vagus nerve towards the brainstem.
	\item \textbf{Signal Integration in the Brainstem}: Upon reaching the NTS, the action potentials induce the release of neurotransmitters. These neurotransmitters interact with downstream neurons, influencing physiological responses such as mood regulation, autonomic functions, and stress responses. 
\end{enumerate}

\begin{figure}[h]
	\centering
	\includegraphics[width=1\columnwidth]{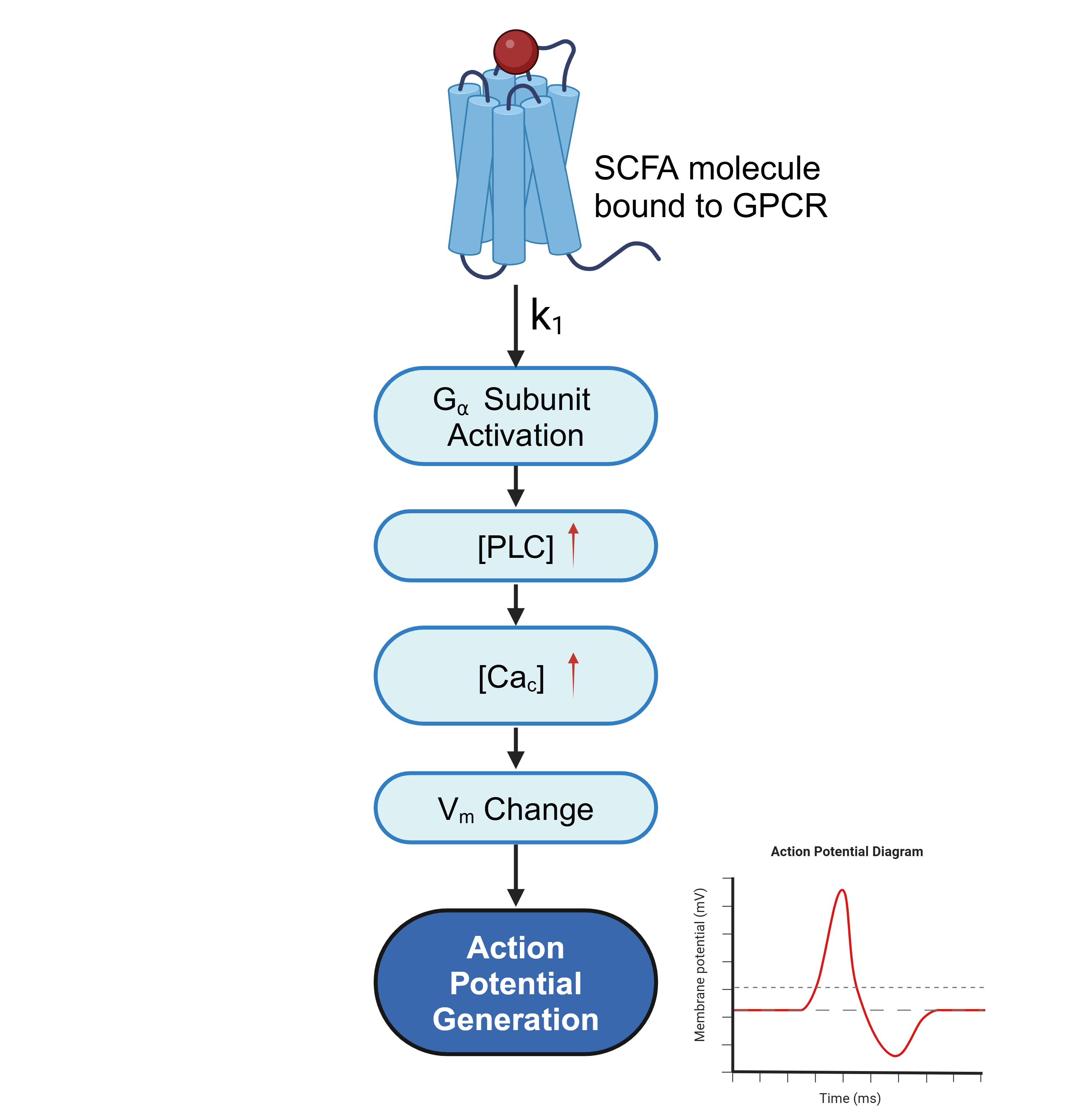}
	\caption{Illustration of ligand-receptor binding.(Created in  https://BioRender.com).}
	\label{fig:LR_illustration}
\end{figure}

The MC-based model explains how microbial metabolites in the gut, specifically SCFAs, influence CNS functions through the vagal pathway. The proposed model allows for the simulation and analysis of signal fidelity, channel capacity, and resilience under physiological conditions by quantifying each stage of the pathway. This comprehensive framework provides an understanding of the mechanisms underlying the GBA, paving the way for future research and therapeutic interventions targeting neurological and psychiatric disorders.

\begin{figure*}[!t]
	\centering
	\includegraphics[width=0.9\textwidth]{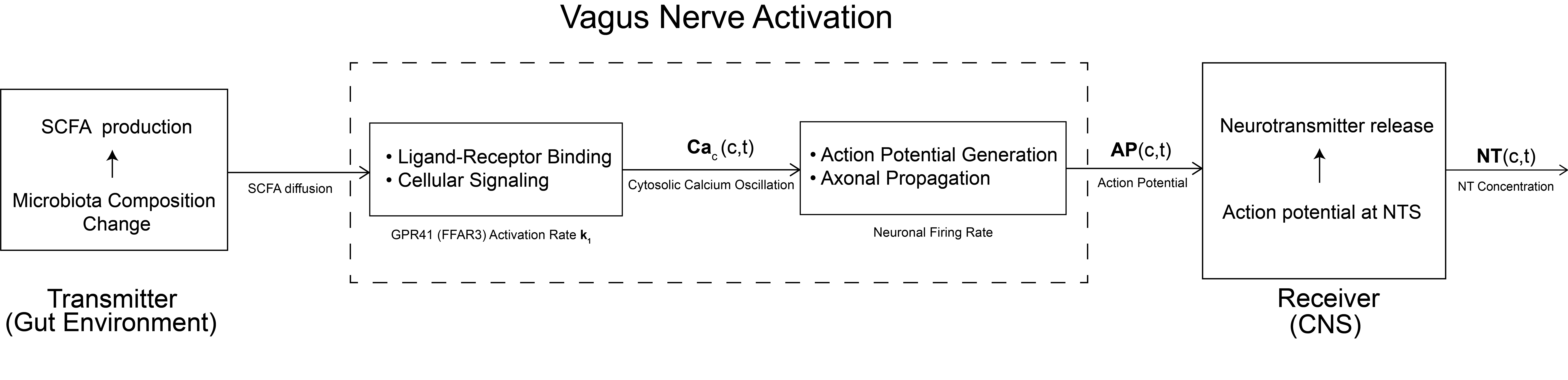}%
	\caption{Scheme of the MC channel model for SCFA-driven vagus nerve signaling withing GBA.}
	\label{fig:vagal_e2e}%
\end{figure*}

As illustrated in the Fig. \ref{fig:LR_illustration}, the G protein alpha subunits (\( [G_\alpha] \)) are activated when SCFAs bind to the FFAR3. This activation triggers a complex intracellular signaling cascade that substantially alters neuronal excitability. The modulation of neuronal excitability through this cascade is crucial as it facilitates the generation of action potentials in vagal afferent fibers. Action potentials are rapid, transient electrical impulses that convey information about the gut environment to the CNS. In this framework, the action potentials generated in response to SCFA binding carry detailed information regarding the composition and activity of the gut microbiota, enabling the vagus nerve to relay subtle signals about the gut environment to the brain.

A key element of this signaling process is the initiation of calcium \((Ca^{2+})\) dynamics. An increase in intracellular calcium concentration triggers depolarization of the neuronal membrane and supports the propagation of action potentials along the vagus nerve. This calcium-mediated mechanism ensures efficient and reliable transmission of gut-related information to the CNS, allowing the CNS to interpret and respond to changes in the gut environment accurately.

Mathematically, the activation of \( [G_\alpha] \) and the following interactions are described by the set of nonlinear ordinary differential equations \cite{kummer2000switching}:

\begin{equation}
    \begin{split}
        \frac{d [G_\alpha]}{dt} &= k_1 + k_2 [G_\alpha] - k_3 [G_\alpha] \frac{[PLC]}{[G_\alpha] + k_4} \\
        &\quad - k_5 [G_\alpha] \frac{[Ca_c]}{[G_\alpha] + k_6} . \\
    \end{split}
\end{equation}

In the governing equation for \([G_\alpha]\), the parameter \( k_1 \) denotes the activation of the GPCR complex as the production rate of G protein alpha subunits induced by SCFA-receptor binding. Adjusting the \( k_1 \) parameter tunes the sensitivity of the pathway and its dynamic range, determining how rapidly the neuron adapts to changing SCFA concentrations. The parameter \( k_2 \) introduces an autoregulatory component by controlling the rate at which \([G_\alpha]\) is maintained, enabling the neuron to sustain a baseline \([G_\alpha]\) level without continuous external input. This self-regulation helps maintain stable signaling and prevents erratic responses as external conditions change.

The parameters \( k_3 \), \( k_4 \), \( k_5 \), and \( k_6 \) govern the interactions between \([G_\alpha]\), \([PLC]\), and cytosolic calcium \([Ca_c]\), emphasizing the role of ER-related processes in the signaling cascade.

To understand how these factors influence the calcium signaling cascade, it is necessary to consider the dynamics of other key molecules. The concentration of \([PLC]\) is governed by:

\begin{equation}
    \frac{d [PLC]}{dt} = k_7 [G_\alpha] - k_8 \frac{[PLC]}{[PLC] + K_9} .\\
\end{equation}

Here, \( k_7 \) governs the synthesis rate of \( [PLC] \), controlling how quickly \( [PLC] \) is produced within the cell. The parameter \( k_8 \) manages the degradation rate of \( [PLC] \), affecting how swiftly \( [PLC] \) is broken down and removed from the system. Additionally, \( k_9 \) describes the Michaelis-Menten constant for \( [PLC] \) degradation, influencing the efficiency of \( [PLC] \) breakdown. \([PLC]\) is synthesized at a rate proportional to \([G_\alpha]\), enabling increased \([G_\alpha]\) levels to upregulate downstream effectors\cite{kummer2000switching}.

Cytosolic calcium concentration \([Ca_c]\) is one of the critical variables driving neuronal excitability since it directly influences the membrane polarization. The dynamics of cytosolic calcium concentration are described by following differential equations:

\begin{equation} 
    \begin{split} 
        \frac{d [Ca_c]}{dt} &= k_{10} [Ca_c] [PLC] \frac{[Ca_{ER}]}{[Ca_{ER}] + k_{11}} + k_{12} [PLC] \\
        &+ k_{13} [G_\alpha] - k_{14} \frac{[Ca_c]}{[Ca_c] + k_{15}} - k_{16} \frac{[Ca_c]}{[Ca_c] + k_{17}} .
    \end{split}
\end{equation}

This formulation emphasizes the role of  \([Ca_c]\) as a dynamically regulated variable that is influenced by multiple factors. Calcium uptake from the endoplasmic reticulum (\([Ca_{ER}]\)) and activation by \([PLC]\) and \([G_\alpha]\) can increase \([Ca_c]\). Conversely, calcium efflux and reuptake mechanisms maintain calcium homeostasis \cite{kummer2000switching}. Depending on the system parameters and initial conditions, these interactions can lead to either oscillatory or steady-state calcium concentration profiles.

The concentration of calcium in the ER (\([Ca_{ER}]\)) is described by:

\begin{equation} 
    \begin{split}
        \frac{d [Ca_{ER}]}{dt} &= -k_{10} [Ca_c] [PLC] \frac{[Ca_{ER}]}{[Ca_{ER}] + k_{11}} \\
        &+ k_{16} \frac{[Ca_c]}{[Ca_c] + k_{17}} ,
    \end{split}
\end{equation}

\noindent where the parameters \( k_{10} \), \( K_{11} \), \( k_{16} \), and \( k_{17} \) are critical for regulating calcium dynamics through the ER. Specifically, \( k_{10} \) controls the uptake rate of calcium into the ER, \( k_{11} \) determines the affinity for calcium release from the ER, \( k_{16} \) manages the reuptake rate of calcium back into the ER, and \( k_{17} \) sets the sensitivity of calcium reuptake into the ER \cite{kummer2000switching}. These ER-related parameters are important in maintaining calcium homeostasis, combining calcium signaling with the overall pathway, and influencing neuronal excitability and signal stability. The model highlights their essential role in regulating \([Ca_c]\) dynamics, ensuring proper neuronal function and responsiveness to external stimuli. Collectively, these parameters ensure the precise regulation of calcium signaling within the cell, affecting everything from signal initiation and propagation to termination and cellular responsiveness.

\subsection{Neuro-spike Communication}

Intracellular calcium oscillations are a key determinant of neuronal excitability and information encoding. Elevated \([Ca_c]\) enhance the probability of action potential generation in vagal afferent fibers. Action potentials serve as the fundamental signals in neural communication, conveying information about gut microbiota activity to the CNS. This calcium-mediated process ensures that subtle changes in SCFA concentrations can be efficiently transformed into electrochemical signals propagating along the vagus nerve.

The parameter \( k_1 \) manages how promptly and robustly the system responds to changes in SCFA levels and directly affects the capacity of the system to carry information. A higher \( k_1 \) value allows the neuron to detect small shifts in SCFA concentration more sensitively, enhancing channel capacity and improving signal fidelity. In contrast, a lower \( k_1 \) value reduces the sensitivity, affecting the throughput and adaptiveness of the signaling pathway.

Upon activation of the calcium signaling pathway, significant changes occur in the neuronal membrane potential that further leads to the generation of action potentials. Elevated  \([Ca_c]\) enhances neuronal excitability by influencing various ion channels embedded in the neuronal membrane. The neuronal membrane potential (\( V_m \)) is governed by the balance of ionic currents flowing through different ion channels and capacitive currents. To analyze these processes, the Hodgkin-Huxley (HH) model \cite{hausser2000hodgkin} provides a comprehensive mathematical framework that describes the electrical behavior of excitable cells, such as neurons. The HH model characterizes the membrane potential \( V_m \) through differential equations. The fundamental equation governing the membrane potential is:

\begin{equation}
C_m \frac{dV}{dt} = I_{\text{ext}} - I_{\text{Na}} - I_{\text{K}} - I_{\text{L}} - I_{\text{CaK}}, \label{eq:V_repeat}
\end{equation}

\noindent where \( C_m \) is the membrane capacitance, \( I_{\text{ext}} \) is an externally applied current, and \( I_{\text{Na}}, I_{\text{K}}, I_{\text{L}}, I_{\text{CaK}} \) represent the sodium, potassium, leak, and calcium-activated potassium currents, respectively. These ionic currents, which are determined by the conductances and voltages of their respective channels, control the flow of ions across the membrane and thus shape the membrane potential.

Each ionic current is defined by its conductance, gating variables, and driving force, as given below:

\begin{align}
I_{\text{Na}} &= g_{\text{Na}} m^3 h (V - E_{\text{Na}}), \label{eq:INa_repeat} \\
I_{\text{K}} &= g_{\text{K}} n^4 (V - E_{\text{K}}), \label{eq:IK_repeat} \\
I_{\text{L}} &= g_{\text{L}} (V - E_{\text{L}}), \label{eq:IL_repeat} \\
I_{\text{CaK}} &= g_{\text{CaK}} [Ca_c] (V - E_{\text{CaK}}). \label{eq:ICaK_repeat}
\end{align}

Here, \( g_{\text{Na}} \), \( g_{\text{K}} \), \( g_{\text{L}} \), and \( g_{\text{CaK}} \) are the maximum conductances for the respective ion channels, while \( E_{\text{Na}} \), \( E_{\text{K}} \), \( E_{\text{L}} \), and \( E_{\text{CaK}} \) are their corresponding reversal potentials. The gating variables \( m \), \( h \), and \( n \) define the probability that ion channels are open:

\begin{align}
\frac{dm}{dt} &= \alpha_m(V)(1 - m) - \beta_m(V)m, \label{eq:dm_repeat} \\
\frac{dh}{dt} &= \alpha_h(V)(1 - h) - \beta_h(V)h, \label{eq:dh_repeat} \\
\frac{dn}{dt} &= \alpha_n(V)(1 - n) - \beta_n(V)n. \label{eq:dn_repeat}
\end{align}

\begin{figure*}[!t]
	\centering
	\includegraphics[width=1\textwidth]{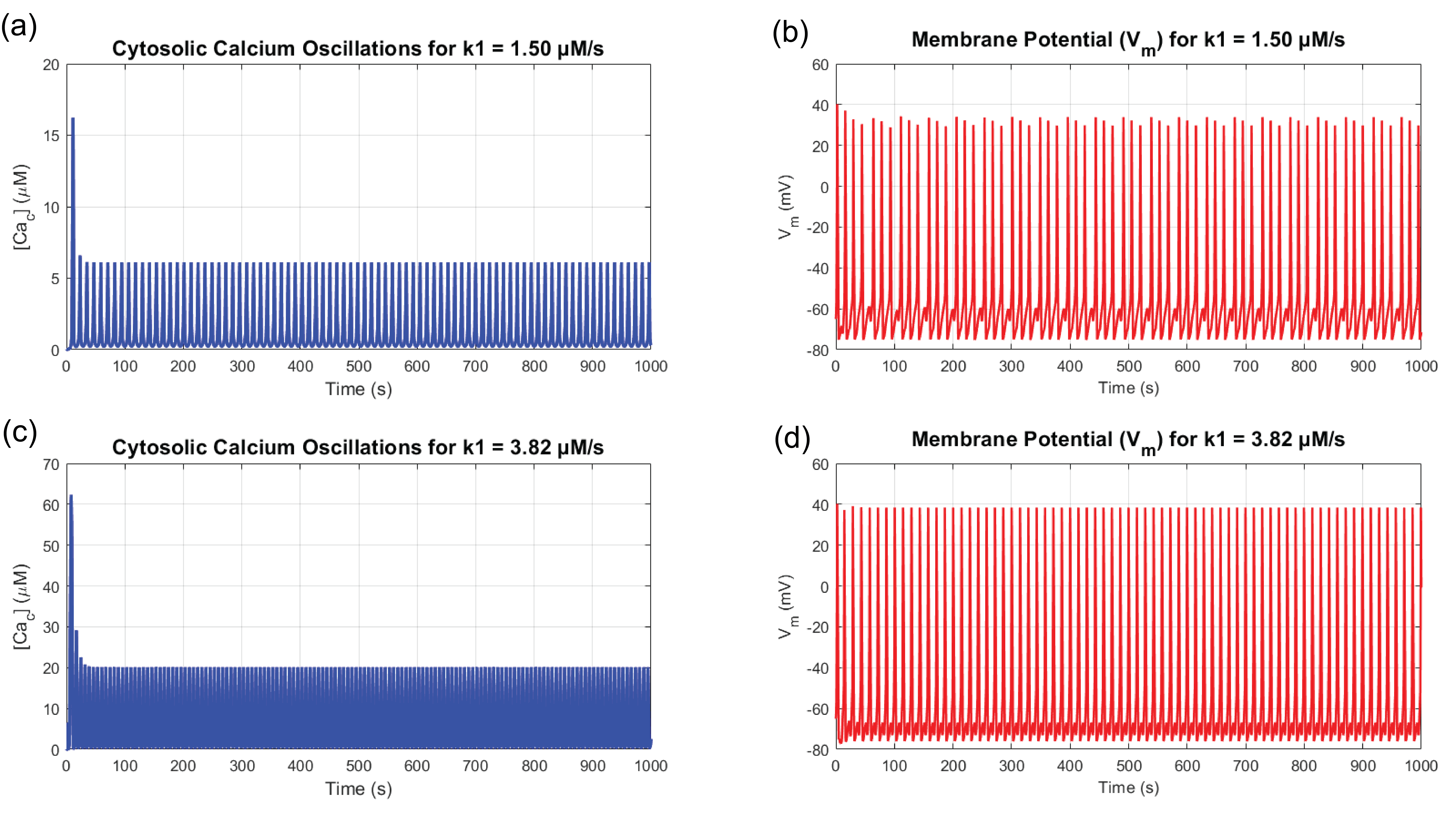}%
	\caption{SCFA-driven cytosolic calcium oscillation and membrane potential for changing $k_1$ values.(a) Cytosolic calcium oscillations for $k_1=1.50 \mu M/s$, (b) Membrane potential $(V_m)$ for $k_1=1.50 \mu M/s$, (c) Cytosolic calcium oscillations for $k_1=3.82 \mu M/s$, (b) Membrane potential $(V_m)$ for $k_1=3.82 \mu M/s$,}
	\label{fig:ca_vm}
\end{figure*}

The rate constants \( \alpha \) and \( \beta \) for each gating variable are voltage-dependent, capturing the kinetics of channel opening and closing in response to changes in membrane potential:

\begin{equation}
    \begin{split}
    \alpha_m(V) &= \begin{cases}
    1.0, & V = -40 \text{ mV} \\
    \frac{0.1(V + 40)}{1 - e^{-(V + 40)/10}}, & \text{otherwise}
        \end{cases}  \\ \\
    \beta_m(V) &= 4.0 \cdot e^{-(V + 65)/18}   \\
\end{split}
\end{equation}

\begin{equation}
    \begin{split}
        \alpha_h(V) &= 0.07 \cdot e^{-(V + 65)/20} \\
        \beta_h(V) &= \frac{1}{1 + e^{-(V + 35)/10}} \\
    \end{split}
\end{equation}

\begin{equation}
    \begin{split}
        \alpha_n(V) &= \begin{cases}
         0.1, & V = -55 \text{ mV} \\
        \frac{0.01(V + 55)}{1 - e^{-(V + 55)/10}}, & \text{otherwise}
        \end{cases} \\ \\
        \beta_n(V) &= 0.125 \cdot e^{-(V + 65)/80}
    \end{split}
\end{equation}

These voltage-dependent rate constants ensure that the gating variables accurately track changes in membrane potential, allowing ion channels to open and close at appropriate time intervals. Through this mechanism, the neuron can reliably initiate and propagate action potentials in response to voltage fluctuations \cite{hausser2000hodgkin}.

As ionic currents and gating variables interact, they collectively shape the temporal evolution of the membrane potential. Elevated \([Ca_c]\) levels, in particular, influence the calcium-activated potassium current \((I_{\text{CaK}})\), enhancing the repolarization phase of the action potential. This adjustment shortens the action potential duration and reinforces the refractory period, ensuring unidirectional propagation and maintaining precise timing and firing patterns. Through these mechanisms, essential physiological information, including signals related to gut microbial activity, is encoded into coherent electrical patterns suitable for transmission.

Table~\ref{tab:hh_params} provides the Hodgkin-Huxley parameters employed in the model. The Hodgkin-Huxley framework describes the electrical characteristics of excitable cells, such as neurons, by incorporating the dynamics of voltage-gated ion channels. Integrating the Hodgkin-Huxley equations with the SCFA-driven signaling ODEs enables the development of a unified framework that links molecular input signals (SCFA concentrations) to electrical outputs (action potentials) within the vagal pathway.

\begin{table}[h]
\caption{Hodgkin-Huxley Model Parameters}
\label{tab:hh_params}
\centering
\begin{tabular}{@{}lll@{}}
\toprule
\textbf{Parameter} & \textbf{Description} & \textbf{Value} \\ \midrule
$C_m$ & Membrane capacitance & $1.0$ µF/cm$^2$ \\
$g_{\text{Na}}$ & Maximum sodium conductance & $120.0$ mS/cm$^2$ \\
$g_{\text{K}}$ & Maximum potassium conductance & $36.0$ mS/cm$^2$ \\
$g_{\text{L}}$ & Leak conductance & $0.3$ mS/cm$^2$ \\
$g_{\text{CaK}}$ & Calcium-activated potassium conductance & $0.1$ mS/cm$^2$ \\
$E_{\text{Na}}$ & Sodium reversal potential & $50.0$ mV \\
$E_{\text{K}}$ & Potassium reversal potential & $-77.0$ mV \\
$E_{\text{L}}$ & Leak reversal potential & $-54.387$ mV \\
$E_{\text{CaK}}$ & Calcium-activated potassium reversal potential & $-80.0$ mV \\
$I_{\text{ext}}$ & External current & $10.0$ µA/cm$^2$ \\
\bottomrule
\end{tabular}
\end{table}

The parameter \( k_1 \) plays a pivotal role in regulating the production rate of G protein alpha subunits \([G_\alpha]\) in response to varying SCFA concentrations. By controlling GPCR activation, \( k_1 \) directly influences phospholipase C \([PLC]\) activation and the subsequent release of calcium from the endoplasmic reticulum \([Ca_{ER}]\). As depicted in Fig.~\ref{fig:ca_vm}, increasing \( k_1 \) results in more frequent cytosolic calcium oscillations, comparing Fig.~\ref{fig:ca_vm}(a) with Fig.~\ref{fig:ca_vm}(c). These enhanced oscillations elevate neuronal excitability and lead to improved, more precisely shaped action potential profiles, as illustrated in the membrane potential plots of Fig.~\ref{fig:ca_vm}(b)-(d). While the total number of action potentials increases with respect to higher \( k_1 \), the increment remains relatively modest, reflecting the intrinsic firing constraints of vagal afferent neurons. Overall, changes in \( k_1 \) modulate the responsiveness of the neuron and its capacity to transmit information effectively through action potentials, which further enhance the ability of the vagus nerve to convey detailed information about the gut microbiota to the CNS.

Neurotransmitter release occurs when the membrane potential $V_m$ reaches a threshold value of $V_{\text{th}} = 20 \,\text{mV}$. This process promotes synaptic communication and ultimately allows the neuron to influence downstream targets. To capture these dynamics, the neurotransmitter concentration \([NT]\) in the synaptic cleft is modeled to incorporate both passive decay and active release triggered by neuronal firing. The neurotransmitter concentration evolves according to:

\begin{equation}
\frac{d[NT]}{dt} = -\frac{[NT]}{\tau_{\text{rec}}}, \label{eq:T_repeat}
\end{equation}

\noindent where \( \tau_{\text{rec}} \) is the neurotransmitter reuptake constant, representing the rate of neurotransmitter removal from the synaptic cleft to terminate the signal. The reuptake process ensures that the released neurotransmitter levels remain controlled, preventing continuous stimulation and helping to maintain signal fidelity. As a result, the interplay between action potential generation, neurotransmitter release, and reuptake provides a clear and physiologically relevant representation of how molecular signals, once translated into electrical impulses, drive synaptic transmission and influence neural circuits within the CNS.

Neurotransmitter release is modeled as a stochastic process that accounts the multivesicular release process \cite{yu2022modeling,ramezani2017information}. The number of vesicles  released per action potential, \( k \), follows a binomial distribution defined as:

\begin{equation}
k \sim \text{Binomial}(N, p_{\text{nt}}), \label{eq:vesicle_release_repeat}
\end{equation}

\noindent where \( N \) represents the total number of the vesicles within the presynaptic neuron, and \( p_{\text{nt}} \) is the probability of vesicle release per action potential. The instantaneous increase in neurotransmitter concentration resulting from the release of \( k \) vesicles is given by:

\begin{equation}
\Delta [NT] = \frac{k \cdot NT_{\text{ves}}}{V_{\text{syn}}}, \label{eq:deltaT_repeat}
\end{equation}

\noindent where \( NT_{\text{ves}} \) represents the neurotransmitter amount of a vesicle and \( V_{\text{syn}} \) denotes the synaptic cleft volume. This formulation ensures that each vesicle release event delivers a definite and measured amount of neurotransmitter to the synaptic cleft, allowing precise and well-regulated synaptic signaling \cite{yu2022modeling}.

\begin{figure*}[!t]
	\centering
	\includegraphics[width=1\textwidth]{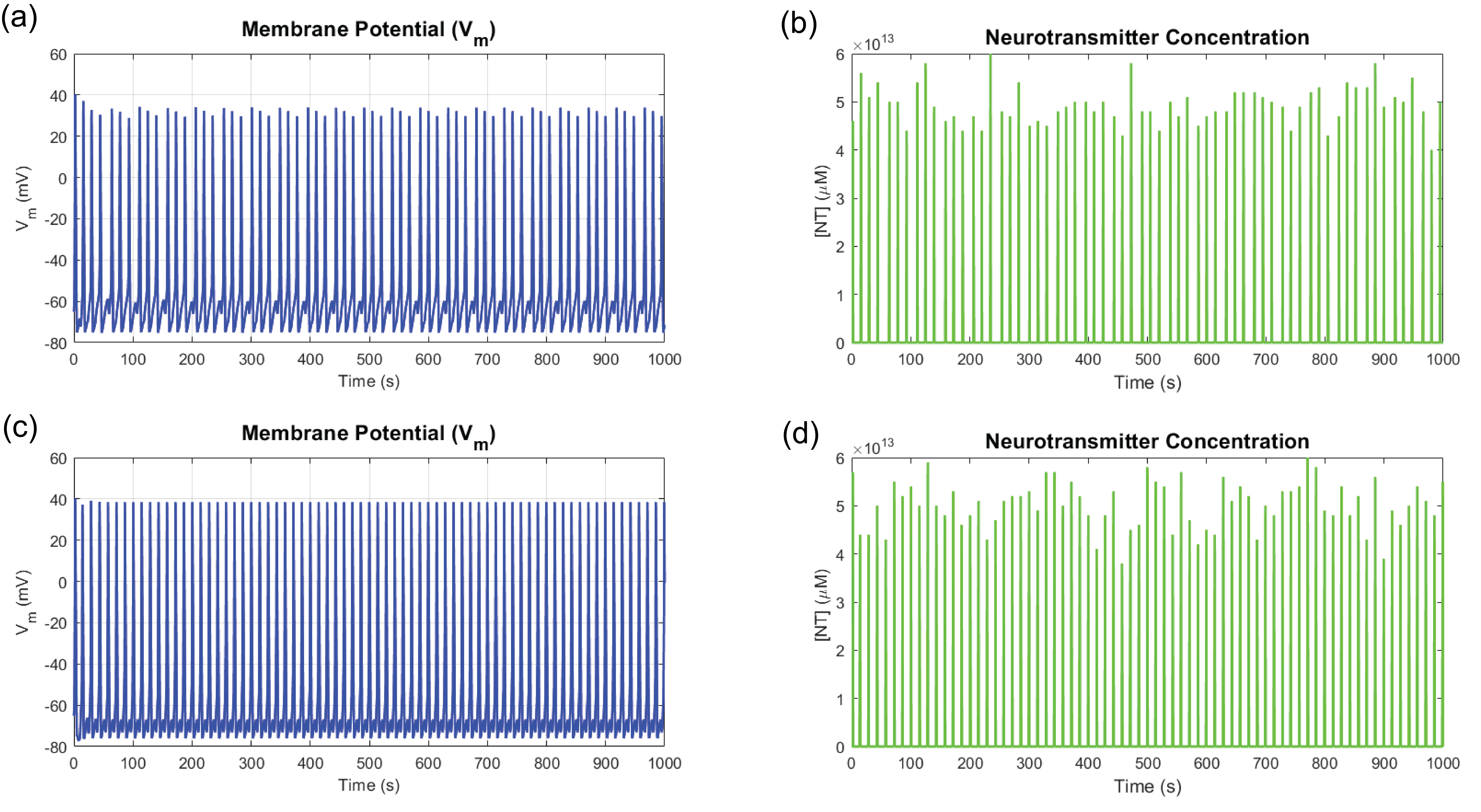}%
	\caption{Action potential generation and neurotransmitter release w.r.t changing activation rate.}
	\label{fig:vm_nt}
\end{figure*}

As illustrated in Fig.~\ref{fig:vm_nt}, the increase in action potential frequency with respect to higher GPCR activation leads to more frequent vesicle release events and thus elevates \([NT]\) levels in the synaptic cleft. By combining passive decay with probabilistic release, this model captures the temporal dynamics of \([NT]\) in the synaptic cleft. High neuronal activity, leading to frequent action potentials, increases vesicle release, raising \([NT]\) and thus enhancing synaptic transmission. Conversely, reuptake processes restore \([NT]\) to baseline, preventing overstimulation and ensuring reliable communication.

Table~\ref{tab:nt_params} summarizes the key parameters governing neurotransmitter release. Adjusting these parameters allows the model to replicate various synaptic behaviors, providing insight into the principles behind learning, memory, and synaptic plasticity.  

\begin{table}[h]
\caption{Neurotransmitter Release Parameters}
\label{tab:nt_params}
\centering
\small
\begin{tabular}{@{}ll@{}}
\toprule
\textbf{Parameter} & \textbf{Description and Value} \\ \midrule
$N$ & Total vesicles: $100$ \\
$NT_{\text{ves}}$ & Neurotransmitter per vesicle: $1 \times 10^{-6}$ mol \\
$V_{\text{syn}}$ & Synaptic cleft volume: $1 \times 10^{-18}$ cm$^3$ \\
$p_{\text{mvr}}$ & Vesicle release probability: $0.5$ \\
$\tau_{\text{rec}}$ & Reuptake time constant: $10$ ms \\
$V_{\text{th}}$ & Action potential threshold: $20$ mV \\
\bottomrule
\end{tabular}
\end{table}

The combined channel modeling presented in this work demonstrates a direct connection between gut-derived molecular signals and the electrical and synaptic activity of vagal neurons. By integrating the SCFA-induced calcium oscillation model, the Hodgkin-Huxley-based action potential generation, and the probabilistic neurotransmitter release dynamics, this end-to-end framework provides a comprehensive representation of SCFA-driven vagal communication throughout the GBA.

\section{Theoretical Analysis and Simulations} 

In this section, the end-to-end MC model derived for SCFA-driven vagus nerve signaling in the gut-brain axis is evaluated using MATLAB-based numerical simulations. The SCFA-induced intracellular biochemical processes, including G protein activation and calcium-mediated cascades, are incorporated into the model to generate electrical action potentials and subsequent neurotransmitter release events at the synaptic terminal.

ICT metrics are employed to examine the overall efficacy of this communication channel. The mutual information between SCFA-induced calcium oscillations and neurotransmitter release events quantifies the information shared between these two processes, providing insight into how effectively information is transmitted through the neuronal channel\cite{civas2020rate,ramezani2018information}. This metric captures the dependency between calcium signaling, which encodes information about gut microbial activity, and the resultant neurotransmitter release that propagates signals to the CNS. Concurrently, delay analysis is conducted to assess the temporal latency between calcium oscillations and neurotransmitter release, providing critical insights into the responsiveness and synchronization of the signaling pathways. By integrating these ICT metrics, we can elucidate the efficiency and reliability of the end-to-end signaling process, highlighting potential bottlenecks and optimizing parameters for enhanced communication fidelity.

The simulations are conducted under varying G-protein activation rates (\( k_1 \)) to understand better the influence of intracellular signaling parameters on the information transfer characteristics and temporal precision of the channel. Specifically, Monte Carlo simulations are conducted by sampling the SCFA-driven GPCR activation rate from independent lognormal distributions. The means of these distributions are set to a range of median values $(1.82, 2.25, 2.68, 3.10, 3.67)$, and each distribution maintained the same mean-to-standard deviation ratio. The system parameters, initial conditions, and simulation details remain consistent with those outlined in the preceding sections. In addition to confirming the fundamental relationship between SCFAs and vagal neuronal activity, the analyses establish a rigorous framework for understanding the end-to-end information transfer within a complex biological network. Applying ICT principles provides a more precise viewpoint on optimizing communication parameters and informing therapeutic interventions that target gut-brain signaling pathways. This approach contributes to a more systematic understanding of the molecular and cellular communication processes underlying neurological function and behavior, and it highlights the value of information-theoretic methodologies in advancing the analysis of biological signaling systems.

\subsection{Mutual Information}

The mutual information analysis aims to quantify the statistical dependence between SCFA-induced calcium oscillations and neurotransmitter release events and serves as a fundamental metric for assessing the information transmission capacity of the neuronal communication channel. Mutual information enables the measuring of the amount of information shared between two random variables, in this case, the occurrence of calcium peaks (\( X \)) via random GPCR activation rate and the subsequent neurotransmitter release events (\( Y \)). This metric effectively measures how much knowing the state of one variable reduces the uncertainty of the other, thereby reflecting the ability of the channel to convey information.

To compute mutual information, binary event sequences are generated by discretizing the simulation time into fixed intervals. Specifically, the simulated time series is first divided into uniform time bins (e.g., \(\Delta t = 1\,\text{s}\)). We generate two binary sequences for each time bin: \( X \) for calcium events and \( Y \) for neurotransmitter release events. An element in these sequences is assigned to \( 1 \) if a corresponding event occurs within the time bin and \( 0 \) otherwise. This technique reduces complex, continuous data into discrete random variables suitable for information-theoretic computations.

After the data is discretized, the joint and marginal probabilities of the events are calculated. The joint probability \( P_{XY}(x, y) \) represents the likelihood of observing a specific combination of calcium and neurotransmitter events within the same time bin. Marginal probabilities \( P_X(x) \) and \( P_Y(y) \) are derived by summing the joint probabilities over the possible states of the other variable. The mutual information \( I(X; Y) \) is computed as:

\begin{equation}
I(X; Y) = \sum_{x=0}^1 \sum_{y=0}^1 P_{XY}(x, y) \log_2 \left( \frac{P_{XY}(x, y)}{P_X(x) P_Y(y)} \right). \label{eq:mutual_info_repeat}
\end{equation}

The resulting mutual information value, measured in bits, quantifies the information transmission capacity of the neuronal channel. Higher mutual information values indicate a more efficient communication process, where intracellular calcium signaling reliably predicts following synaptic neurotransmitter release events. In the presented simulations, mutual information is computed for different values of \( k_1 \) to compare how altered G-protein activation rates affect the information-carrying capacity of the defined communication channel.

\begin{figure}[h]
	\centering
	\includegraphics[width=1\columnwidth]{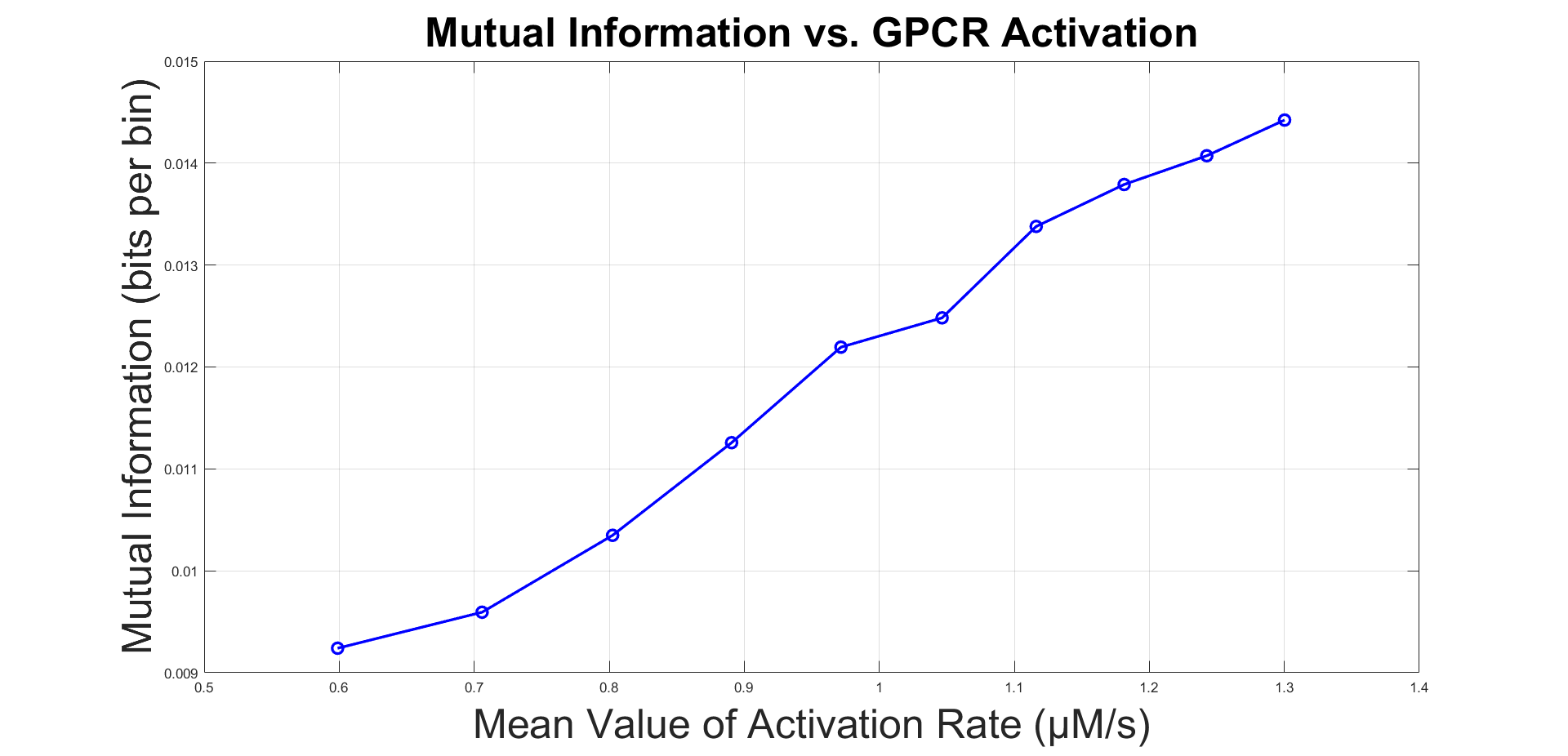}
	\caption{Mutual information vs GPCR activation rate.}
	\label{fig:mutual_vs_k1_monte}
\end{figure}

As illustrated in Fig. \ref{fig:mutual_vs_k1_monte}, the simulation results indicate that increasing the G-protein activation rate \( k_1 \) leads to a noticeable improvement in mutual information. The activation rate \( k_1 \) depends on the type of bound SCFA and ranges between \(1.52\,\mu\text{M/s}\) and \(3.82\,\mu\text{M/s}\) \cite{kummer2000switching}. This variability arises from the inherently random nature of the binding process in the gut and is effectively modeled using a lognormal distribution \cite{furusawa2005ubiquity}. Specifically, for \( k_1 = 1.52\,\mu\text{M/s} \), the mutual information was found to be \(0.0085\) bits per bin, while raising \( k_1 \) to \( 3.82\,\mu\text{M/s} \) increased the mutual information to \(0.0110\) bits per bin. Although these mutual information values are relatively modest, the enhancement observed with a higher \( k_1 \) suggests a more effective coupling between intracellular calcium signaling and synaptic neurotransmitter release.

The observed increase in mutual information aligns with the theoretical expectation that strengthening the intracellular signaling cascade leads to more reliable information encoding. As G-protein-mediated responses become more evident, the timing and occurrence of calcium oscillations gain greater predictive power regarding the subsequent release of neurotransmitters. This improved correlation implies that the channel can better convey subtle variations in SCFA-induced biochemical activity, thereby serving as a more faithful representation of underlying gut-brain communication processes.

\begin{figure}[h]
	\centering
	\includegraphics[width=1\columnwidth]{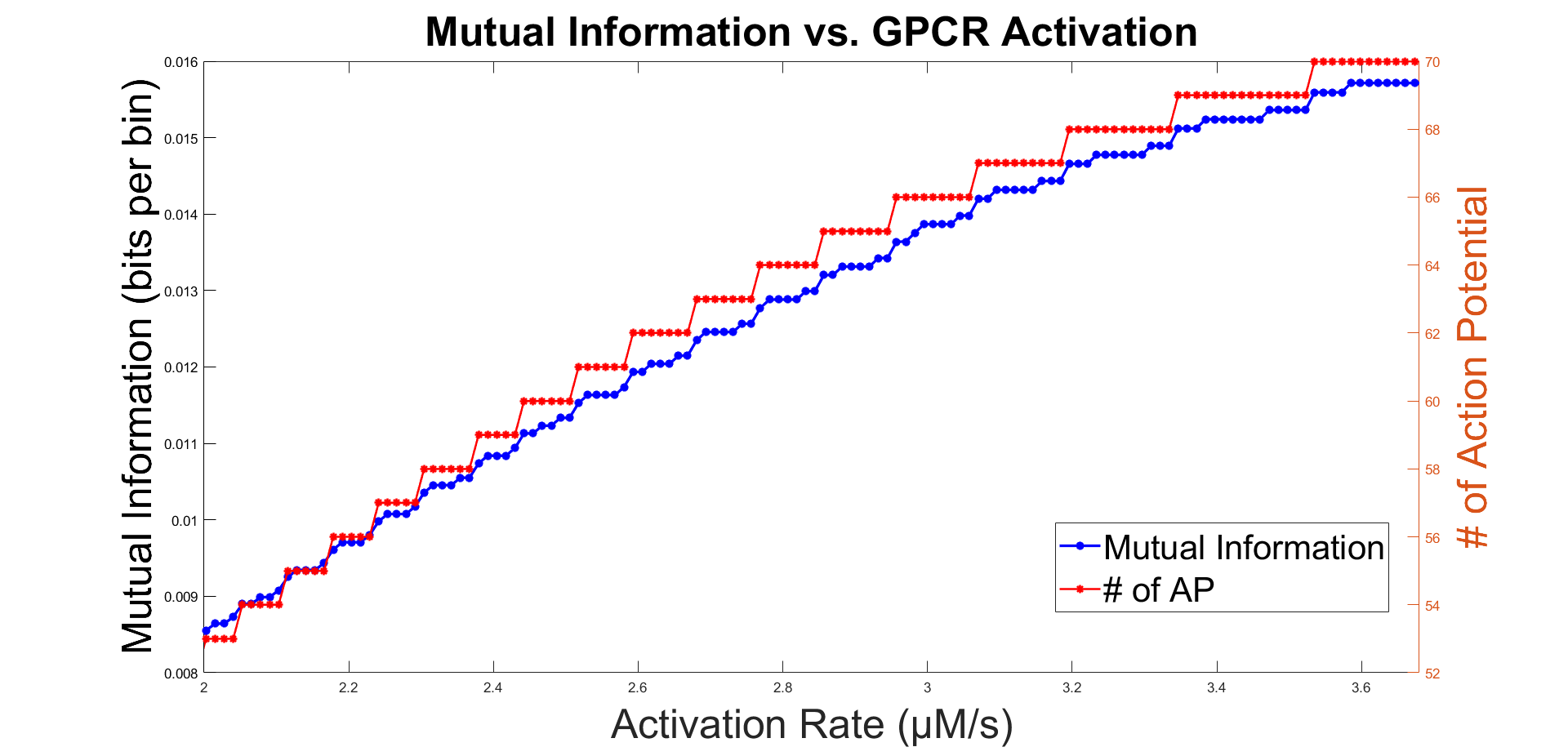}
	\caption{Action potential and mutual information vs GPCR activation rate.}
	\label{fig:mutual_vs_k1_ap}
\end{figure}

A higher GPCR activation rate also overlaps with increased action potential frequency, driving more frequent vesicle release events. As shown in Fig. \ref{fig:mutual_vs_k1_ap}, this increased neuronal firing contributes to the rise in mutual information, as the tighter temporal alignment between calcium oscillations and action potentials augments the ability of the channel to capture subtle variations in SCFA-induced activity. Consequently, strengthening the GPCR-mediated intracellular cascade not only refines the timing of neurotransmission but also increases the overall fidelity of gut-brain communication.

The obtained results align with the theoretical hypothesis that intensifying the intracellular signaling network improves information encoding. As G-protein-mediated responses become more prominent, the timing and amplitude of calcium oscillations more accurately cause neurotransmitter release, thereby expanding the capacity of the system to represent essential biochemical signals. From a broader perspective, the sensitivity of mutual information to GPCR activation rate highlights the feasibility of tuning intracellular parameters to optimize gut-brain signaling. Such insights may guide therapeutic interventions aimed at modulating GPCR activity and enhancing neural communication under pathological conditions, thus underlining the value of information-theoretic approaches in understanding complex biological networks.

\subsection{Delay Analysis}

While mutual information provides insights into the informational coupling between input and output events, delay analysis provides an important measure of the temporal dynamics inherent in the SCFA-mediated signaling pathway\cite{akan2016fundamentals}. Specifically, it quantifies the latency between cytosolic calcium peaks and the following neurotransmitter release events. This delay designates the time required for intracellular calcium signaling to culminate in neurotransmitter release, thereby reflecting the responsiveness and synchronization of the neuronal communication channels.

The delay is computed by identifying each calcium peak within the simulation data and determining the time of the earliest neurotransmitter release event that occurs at or after that peak. The delay \( \Delta t_i \) for the \( i \)-th calcium peak is defined as:

\begin{equation}
\Delta t_i = t_{\text{NT}, j} - t_{\text{Ca}, i}, \label{eq:delay_repeat}
\end{equation}

\noindent where \( t_{\text{Ca}, i} \) is the time of the \( i \)-th calcium peak, and \( t_{\text{NT}, j} \) is the time of the earliest neurotransmitter release event observed at or following \( t_{\text{Ca}, i} \). A distribution of delay values that characterize the temporal relationship between calcium signaling and neurotransmitter release is obtained by performing this calculation for all detected calcium peaks.

To summarize the delay characteristics, the mean of delay \( \mu_{\Delta t} \) and the standard deviation of the delay \( \sigma_{\Delta t} \) are computed as:

\begin{align}
\mu_{\Delta t} &= \frac{1}{N_{\text{delays}}} \sum_{i=1}^{N_{\text{delays}}} \Delta t_i, \label{eq:mean_delay} \\
\sigma_{\Delta t} &= \sqrt{\frac{1}{N_{\text{delays}} - 1} \sum_{i=1}^{N_{\text{delays}}} (\Delta t_i - \mu_{\Delta t})^2}. \label{eq:std_delay}
\end{align}

Here, \( \mu_{\Delta t} \) represents the average latency between calcium peaks and neurotransmitter release events, while \( \sigma_{\Delta t} \) quantifies the variability in these intervals. A lower mean delay indicates that information is transmitted more rapidly. Conversely, a higher standard deviation suggests more significant variability in response times, potentially affecting the reliability and synchronization of the signaling pathway.

\begin{figure}[h]
	\centering
	\includegraphics[width=1\columnwidth]{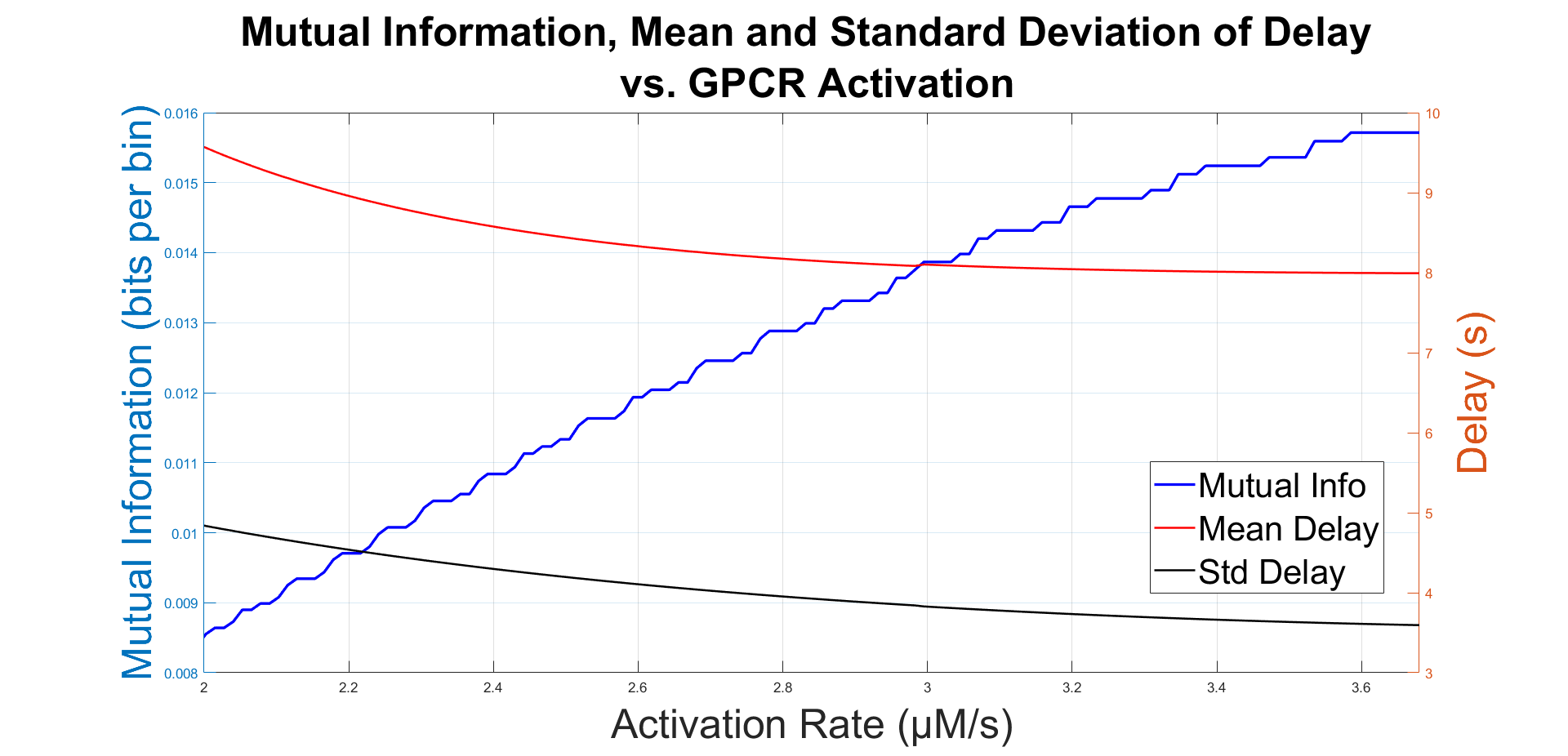}
	\caption{Delay vs GPCR activation rate.}
	\label{fig:delay_k1}
\end{figure}

Fig. \ref{fig:delay_k1} demonstrates that the mean and standard deviation of the delay both decrease as the GPCR activation rate, \(k_1\), increases, reflecting more rapid and consistent signal propagation in SCFA-driven vagus nerve signaling. Since \(k_1\) governs the SCFA-GPCR binding process and the following intracellular cascades, higher values of \(k_1\) accelerate the calcium release and action potential generation processes. As a result, the mutual information depicted in the same figure also rises with increasing \(k_1\), while the delay metrics drop, consistent with theoretical expectations that strengthening GPCR-mediated pathways enhances both the information transfer rate and its temporal fidelity. For example, increasing \(k_1\) from \(1.52\,\mu\text{M/s}\) to \(3.82\,\mu\text{M/s}\) lowers the mean delay from about \(9.40\,\text{s}\) to \(7.99\,\text{s}\) and reduces the standard deviation from \(4.44\,\text{s}\) to \(3.57\,\text{s}\). These findings demonstrate the sensitivity of the proposed communication model to an intrinsic parameter (\(k_1\)), which governs both the degree of informational coupling (via mutual information) and the timing of neural responses.

On the other hand, the obtained results confirm the expectation that SCFA-driven molecular communication along the vagus nerve is significantly slower than conventional electrical synaptic signaling. The involvement of secondary messengers, receptor-ligand binding, and downstream biochemical cascades inherently introduces substantial temporal delays. Although increasing \( k_1 \) improves responsiveness, it does not fundamentally alter the time scale dominated by these biochemical processes. Consequently, the signaling pathway remains slow, consistent with the fundamental principles of molecular communication where information transmission typically involves longer time scales due to diffusion and enzymatic reaction kinetics.

Understanding the delay dynamics provides critical insights into the temporal responsiveness and synchronization of the SCFA-mediated signaling pathway, making it possible to determine how rapidly and consistently molecular events are translated into neural outputs. Characterizing the delay distribution and its sensitivity to intracellular parameters, such as \( k_1 \), highlights potential avenues for reducing latency and variability, thereby improving the speed and reliability of gut-brain communication. Beyond these temporal considerations, the complementary examination of mutual information can further elucidate the information-carrying capacity of the pathway. Integrating mutual information and delay metrics offers a comprehensive evaluation of the system's performance, illustrating how changes in intracellular kinetics influence both the fidelity and timing of information transfer. These analyses enable the identification of key factors that either enhance or impede effective information transmission from the gut microbiota to the CNS. This holistic  approach deepens our understanding of molecular communication mechanisms within biological systems, identifies potential areas for optimization, and advances the investigation of the gut-brain axis. Consequently, it provides valuable insights for developing targeted therapeutic treatment techniques and enhancing overall neurological health.

\section{Discussion}

In this study, we developed a novel end-to-end MC framework to model SCFA-driven vagus nerve signaling within the GBA. The model integrates SCFA-receptor binding, intracellular calcium dynamics, Hodgkin-Huxley-based membrane potential equations, and probabilistic neurotransmitter release to capture the complete communication pathway from gut-derived SCFAs to neural activity in the brainstem.

One of the key outcomes is the role of the G-protein activation rate (\(k_1\)), which is influenced by SCFA type and concentration and varies between \(1.8\,\mu\text{M/s}\) and \(3.7\,\mu\text{M/s}\). This rate follows a lognormal distribution, effectively modeling the inherent randomness of the SCFA-receptor binding process. High \(k_1\) values enhance the sensitivity of SCFA-driven signaling and increase the fidelity of neural signal generation by promoting more frequent calcium oscillations and action potentials. However, the action potential rate remains constrained by the biophysical limits of neuronal firing, preventing disproportionate increases despite elevated GPCR activation.

The mutual information analysis reveals that higher SCFA-induced GPCR activation rates improve the information transmission capacity of the communication channel. Enhanced SCFA-mediated intracellular cascades synchronize calcium oscillations with neurotransmitter release events, thereby strengthening the correlation between SCFA concentrations and neural output. This outcome suggests that modulating intracellular signaling strength can enhance the detection and responsiveness to gut conditions, thereby increasing the robustness and precision of gut-to-CNS information flow.

Additionally, delay analysis indicates that increased activation rates reduce both the mean and variability of signaling delays. Nevertheless, SCFA-driven pathways exhibit slower communication rates than conventional electrical synapses due to the molecular nature of ligand-receptor interactions and secondary messenger dynamics. Our model proposes that adjusting enzymatic rates or receptor activation thresholds could mitigate these delays, offering potential strategies to accelerate the signaling process.

Overall, the MC-based framework demonstrates its effectiveness in explaining SCFA-driven vagus nerve signaling mechanisms and underscores the critical role of intracellular parameters in regulating the rate and reliability of gut-brain communication. This comprehensive model paves the way for exploring diverse scenarios, including different microbial compositions, pathological conditions, and pharmacological interventions, thereby guiding the development of targeted treatments to restore optimal neural communication within the GBA.

\section{Conclusion}

This paper has presented an end-to-end molecular communication (MC) framework for short-chain fatty acid (SCFA)-driven vagus nerve signaling within the gut-brain axis (GBA). The model integrates SCFA-receptor binding kinetics, Hodgkin-Huxley-based neuron excitability, and probabilistic neurotransmitter release, providing a unified view of how gut-derived molecular signals are translated into electrical and synaptic events in the brainstem. The MC-based framework offers a robust tool for investigating how subtle variations in SCFA concentrations impact neuronal function and, eventually, influence brain activity by capturing important biochemical and electrophysiological processes.

The numerical simulations highlight that increasing the G-protein activation rate enhances both the frequency of calcium oscillations and the action potential firing rate. The higher mutual information indicates a more robust link between SCFA concentration and neural output. Delay analysis further shows that increasing the activation rate reduces both the mean and standard deviation of the delay, which means the communication link is effectively sped up. Even though SCFA-driven pathways remain slower than conventional electrical synapses due to the intrinsic features of molecular communication, adjusting enzymatic rates or receptor activation thresholds can partially reduce latency, suggesting strategies to accelerate the signaling process.

From an information-theoretic perspective, the increased GPCR (G-protein coupled receptor) activation rate improves the ability of the channel to seize gut-derived signals, thereby enhancing the reliability of signal transfer to the brain. This result emphasizes the potential for developing therapeutic interventions that modify SCFA levels or receptor sensitivities to mitigate neurological and psychiatric conditions associated with gut dysbiosis or impaired gut-brain communication. By carefully analyzing how different intracellular parameters affect both the fidelity and timing of the signaling process, strategies can be developed to refine communication efficiency.

Future investigations can extend these findings by exploring different microbial compositions and disease states marked by altered gut microbiota or encompassing additional biological routes such as endocrine and immune pathways. Through this broader perspective, further progress in understanding and manipulating gut-brain communication holds promise for advancing therapeutic strategies targeting both neurological and psychiatric disorders.

\bibliographystyle{IEEEtran}
\bibliography{References}

\end{document}